\documentclass[%
 reprint,
 amsmath,amssymb,
 aps,
]{revtex4-1}

\usepackage[utf8]{inputenc}
\usepackage{float}
\usepackage{subcaption}
\usepackage{graphicx}
\usepackage{dcolumn}
\usepackage{bm}
\usepackage[colorlinks=true, linkcolor=black, urlcolor=black, menucolor=black, citecolor=blue]{hyperref}

\begin{document}

\preprint{APS/123-QED}

\title{Self-Organization Towards $1/f$ Noise in Deep Neural Networks}

\author{Nicholas Chong Jia Le}
 \email{chong.jiale.nicholas@u.nus.edu}
\affiliation{%
 Department of Physics\\
 National University of Singapore, 117551 Singapore
}%
\author{Feng Ling}%
 \email{fengl@ihpc.a-star.edu.sg}
\affiliation{%
 Complex Systems Group, Systems Science Department\\
 Institute of High Performance Computing, A*STAR, 138632 Singapore
}%
\affiliation{%
 Department of Physics\\
 National University of Singapore, 117551 Singapore
}%

\date{\today}

\begin{abstract}

The presence of $1/f$ noise, also known as pink noise, is a well-established phenomenon in biological neural networks, and is thought to play an important role in information processing in the brain. In this study, we find that such $1/f$ noise is also found in deep neural networks trained on natural language, resembling that of their biological counterparts. Specifically, we trained Long Short-Term Memory (LSTM) networks on the `IMDb' AI benchmark dataset, then measured the neuron activations. The detrended fluctuation analysis (DFA) on the time series of the different neurons demonstrate clear $1/f$ patterns, which is absent in the time series of the inputs to the LSTM. Interestingly, when the neural network is at overcapacity, having more than enough neurons to achieve the learning task, the activation patterns deviate from $1/f$ noise and shifts towards white noise. This is because many of the neurons are not effectively used, showing little fluctuations when fed with input data. We further examine the exponent values in the $1/f$ noise in ``internal" and ``external" activations in the LSTM cell, finding some resemblance in the variations of the exponents in fMRI signals of the human brain. Our findings further supports the hypothesis that $1/f$ noise is a signature of optimal learning. With deep learning models approaching or surpassing humans in certain tasks, and being more ``experimentable'' than their biological counterparts, our study suggests that they are good candidates to understand the fundamental origins of $1/f$ noise.
\end{abstract}

\maketitle


\section{\label{sec:intro}Introduction}

Pink noise, also known as $1/f$ noise, is widely observed in many living systems \cite{Munoz:2018aa}. It refers to the power-law relationship between the power spectral densities and frequencies in a temporal signal, with energy decreasing inversely with frequency ($f^{-\beta}$) over a wide range of frequencies \cite{Bak:1987aa}. In biological systems, such noise has been observed in various signals including heart rate fluctuations \cite{noise:heart_phys, noise:heart_bio}, human gait \cite{Hausdorff_1995,Hausdorff_1997}, human optical cells \cite{noise:vision}, and in activity scans of the human brain \cite{OBYRNE2022820}.

Specifically in human brains, activity signals like electroencephalograph (EEG) and functional magnetic resonance imaging (fMRI) \cite{Linkenkaer-Hansen:2001aa,noise:meg, EEG:principles} have found such patterns. The $1/f$ noise in the human brain presents itself differently in a healthy human brain compared to a brain with neurological conditions and aging \cite{Linkenkaer-Hansen:2005aa,Montez_2009,brain:schizophrenia,Voytek:2015aa}. While the study of this form of brain activity is in its early stages due to the $1/f$ signal being regarded as extraneous noise in the past, it has been proposed that $1/f$ noise in the brain is important in regulating function and serves other cognitive purposes \cite{brain:importance_of_noise}. Numerous studies have measured the scaling exponent $\beta$ of EEGs with similar results \cite{EEGMEG:noise, EEG:noise}. The aggregate results obtained in \cite{EEGMEG:noise} gives a scaling exponent $\beta = 1.33 \pm 0.19$, demonstrating $1/f$ noise. In fMRI studies \cite{brain:overview, fMRI:noise}, the scaling exponents measured in these studies are smaller than in EEGs, with an average scaling exponent of $\beta = 0.84$. This exponent became even smaller when the brain performs tasks, averaging to $\beta = 0.72$ across the brain. 

The first attempt to explain $1/f$ noise was the idea of self-organized criticality (SOC) \cite{Bak:1987aa}, where the simple sandpile model was able to generate such fluctuations in terms of the cascading avalanche of sand. It was later found that the neural avalanche in the brain follows similar power-law distributions, near the critical branching parameter value close to 1 \cite{Beggs:2003aa}, and the power-law exponent value is consistent with the mean-field result of the SOC model \cite{Christensen:1993aa}. Later, more studies looked into the synchronization of the brain \cite{Expert_2010}. Combined with simulation studies \cite{Poil:2012aa,Girardi_Schappo_2016,Lombardi_2017,Aguilar_Vel_zquez_2019}, it has been established that $1/f$ noise is associated with critical synchronization states.

While $1/f$ noise has been extensively studied in the biological brain, the existence of this phenomenon has not been clearly established in artificial neural networks (NN) - machine learning models loosely inspired by their biological counterparts \cite{NN:textbook}. Artificial neural networks have achieved impressive performance in many domains, with the most powerful models consisting of very deep hierarchical architectures with large number of parameters and highly complex structures \cite{He_2016_CVPR,NIPS2017_3f5ee243}. However, the precise learning dynamics and computational principles underlying their success remain poorly understood from a theoretical standpoint. On the flip side, these ``artificial brains'' can be better experimental subjects than their biological counterparts, in the sense that they can be controlled and observed at the lowest level possible, without external confounding factors. They are thus ideal subjects for providing a deeper understanding on $1/f$ noise if it exists within them. In this work, we look for the existence of $1/f$ noise in a common deep learning model used for time series data modelling, known as Long-Short-Term Memory (LSTM) networks \cite{NN:LSTM_original} trained on a real-world dataset.

\section{\label{sec:expt}Methods}

There are many successful deep learning models for different tasks, ranging from ResNet specialized for image recognition \cite{He_2016_CVPR} to attention based models for natural language \cite{NIPS2017_3f5ee243}. However, to investigate the existence of $1/f$ noise, we need a network that is used specifically to model time series data of arbitrary lengths, such as Recurrent Neural Networks (RNNs). 
RNNs are a type of network that preserve the state of an input across a temporal sequence by feeding the outputs of its nodes back into those same nodes. This is as opposed to feedforward networks, where data flows only from layer to layer. By retaining knowledge of previous inputs through this recurrence, RNNs are significantly more adept than simple feedforward networks at processing sequential data.

\begin{figure}
\includegraphics[width=\linewidth]{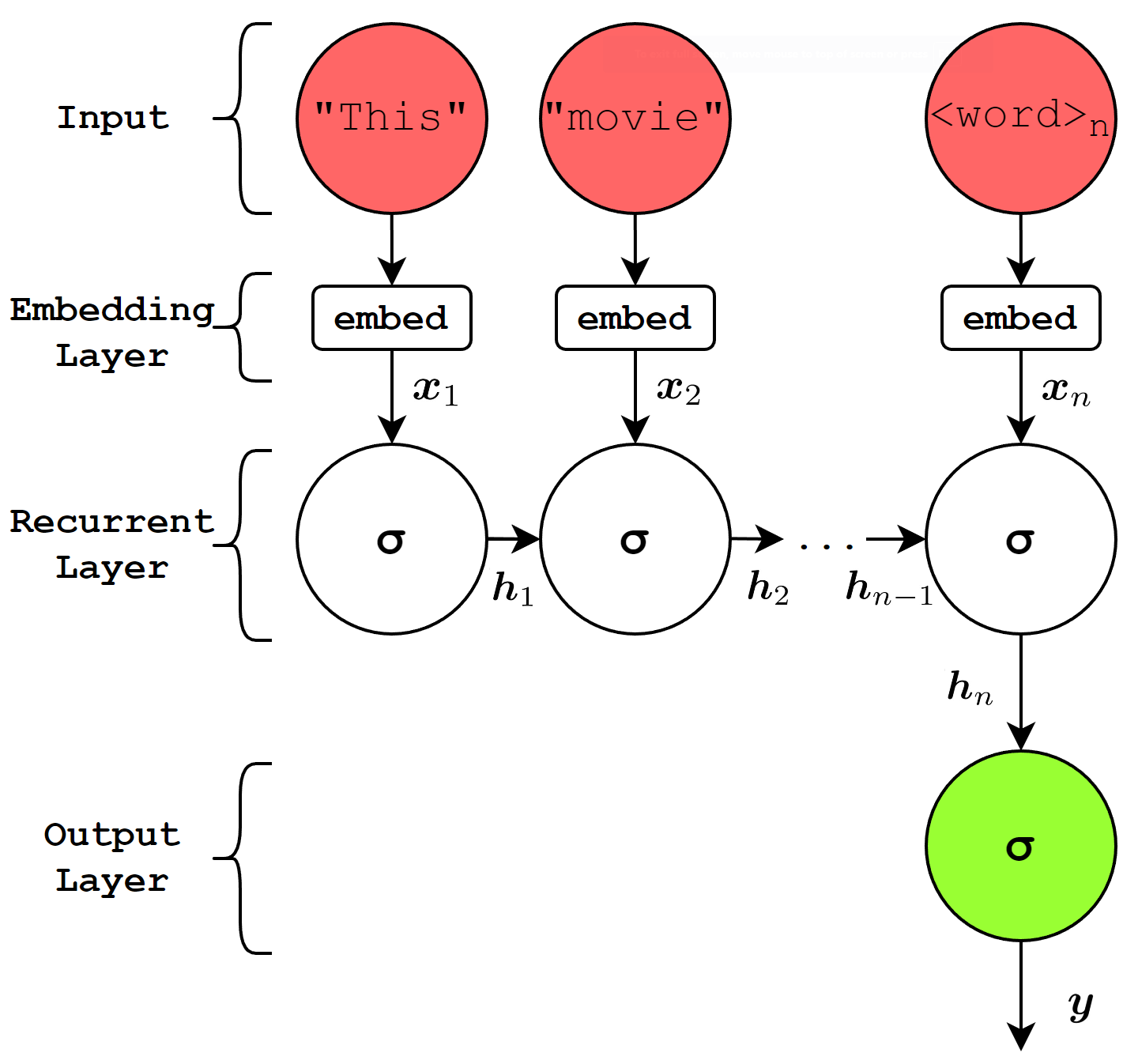}
\caption{\label{fig:rnn_unrolled}A (many-to-one) recurrent neural network visualised in its temporally unrolled representation. A time series (in this case a movie review with $n$ words) is input into the network sequentially. For each time step $t$, the $t$-th word passes into the embedding layer, which converts the word into a vector using a learned representation of a continuous vector space. The vector $\boldsymbol{x}_t$ then passes into the recurrent layer, which accepts both $\boldsymbol{x}_t$ and the output of itself from the previous time step, $\boldsymbol{h}_{t-1}$. The recurrent layer then passes its output, $\boldsymbol{h}_t$, into itself for the next time step. At the final time step $n$, the recurrent layer passes its output $\boldsymbol{h}_n$ to the output layer which converts it to the output $\boldsymbol{y}$.}
\end{figure}

\begin{table*}
\centering
\caption{\label{tab:hyperparam}
Hyperparameters selected for the LSTM networks}
\begin{ruledtabular}
\begin{tabular}{ccccc}
\textrm{Size of embedding layer}&
\textrm{Training batch size}&
\textrm{Dropout factor}&
\textrm{L2 regularisation factor}&
\textrm{Learning rate}\\
\colrule
8 & 128 & 0.1 & 0.001 & 0.005\\
\end{tabular}
\end{ruledtabular}
\end{table*}

\autoref{fig:rnn_unrolled} shows the most basic form of an RNN, demonstrating the idea that these recurrent networks are deep through \textit{time}, in contrast to the depth through \textit{layers} of a simple feedforward network. However, this also means that the simple RNN suffers from the same problem as DNNs with large depths - the vanishing gradient problem. RNNs struggle to converge for particularly long input sequences of more than 10s of time steps. In a way, RNNs are similar to the human brain at an abstract level, as the human brain continuously receives information and processes it using our biological neural networks.

\begin{figure*}   
    \includegraphics[width=\linewidth]{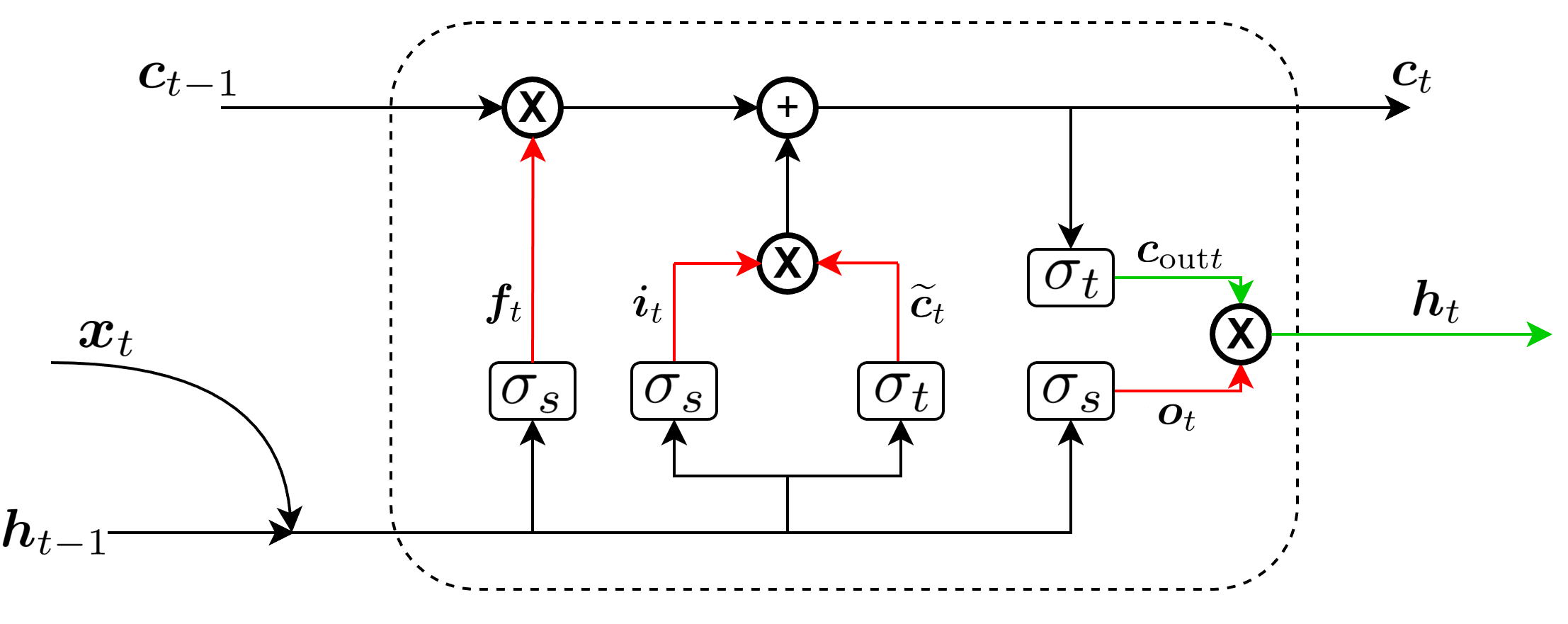}
    \caption{\label{fig:LSTM_highlighted}The LSTM cell (dotted circle) with its internal structure shown. The red lines represent the ``internal" activations while the green lines represent the ``external" activations. $\sigma_t$ and $\sigma_s$ represent the tanh activation and sigmoid activation respectively.}
\end{figure*}

There are many variants of RNNs, ranging from the fully recurrent network, gated recurrent unit network \cite{Cho:2014aa}, bi-directional RNN \cite{Schuster_1997}, to the more commonly used 
reservoir computer \cite{TANAKA2019100} among the physics community. 

In this study we use a particular type of RNN called Long Short-Term Memory (LSTM) networks, due to its outstanding real-world performance \cite{NIPS2014_a14ac55a}. In this architecture, an LSTM cell was created to replace the recurrent cell in the vanilla RNN shown in Figure \ref{fig:rnn_unrolled} in order to solve the vanishing gradient problem \cite{NN:LSTM_original} by maintaining an internal cell state $\boldsymbol{c}$.

An LSTM cell (\autoref{fig:LSTM_highlighted}) which consists of many different activations compared to the single activation in a vanilla RNN cell. This LSTM cell, like the vanilla RNN cell, takes in $\boldsymbol{x}_t$ and $\boldsymbol{h}_{t-1}$ as inputs, along with the additional input of the previous cell state $\boldsymbol{c}_{t-1}$. Like the vanilla RNN, the LSTM cell also outputs the current hidden state $\boldsymbol{h}_t$ and additionally the current cell state $\boldsymbol{c}_t$ to itself in the next time step. The addition of the internal cell state $\boldsymbol{c}$ helps in preserving temporal correlations \cite{NN:LSTM_original}.

The following equations describe the different internal state vectors of an LSTM during a forward pass at time step $t$:

\begin{eqnarray}
    \boldsymbol{i}_t&=&\sigma_s(\boldsymbol{W}_{xi}\boldsymbol{x}_t+\boldsymbol{W}_{hi}\boldsymbol{h}_{t-1}+\boldsymbol{b}_i)
    \\
    \boldsymbol{f}_t&=&\sigma_s(\boldsymbol{W}_{xf}\boldsymbol{x}_t+\boldsymbol{W}_{hf}\boldsymbol{h}_{t-1}+\boldsymbol{b}_f)
    \\
    \boldsymbol{o}_t&=&\sigma_s(\boldsymbol{W}_{xo}\boldsymbol{x}_t+\boldsymbol{W}_{ho}\boldsymbol{h}_{t-1}+\boldsymbol{b}_o)
    \\
    \boldsymbol{\widetilde{c}}_t&=&\sigma_t(\boldsymbol{W}_{xc}\boldsymbol{x}_t+\boldsymbol{W}_{hc}\boldsymbol{h}_{t-1}+\boldsymbol{b}_c)
    \\
    \boldsymbol{c}_t&=&\boldsymbol{f}_t\odot{\boldsymbol{c}_{t-1}}+\boldsymbol{i}_t\odot{\boldsymbol{\widetilde{c}}_t}
    \\
    \boldsymbol{c}_{\mathrm{out}t}&=&\sigma_t(\boldsymbol{c}_t)
    \\
    \boldsymbol{h}_t&=&\boldsymbol{o}_t\odot\boldsymbol{c}_{\mathrm{out}t}
\end{eqnarray}
where $\odot$ is the element-wise product, $\sigma_s$ is the sigmoid activation function $\frac{1}{1+e^{-x}}$, $\sigma_t$ is the $\tanh$ activation function, and $\boldsymbol{W}$ contains the weight matrices of the model.

\subsection{\label{sec:expt-dataset}Dataset}
The dataset chosen for the sentiment analysis task will be the Large Movie Review Dataset \cite{dataset} which consists of 50000 polarised movie reviews obtained from the Internet Movie Database (IMDb) \footnote{https://www.imdb.com/}. This dataset consists of 25000 positive reviews (score $\ge$ 7 out of 10) and 25000 negative reviews (score $\le$ 4 out of 10).

Preprocessing steps such as the removal of punctuation and converting of words to lowercase were performed. The words were also converted to tokens, with the top 4000 words (88.3\% of the full vocabulary) converted into unique tokens, and the rest of the words converted into a single [UNK] token.

\subsection{\label{sec:expt-lstm}LSTM network architecture}
The LSTM network will consist of three layers: An embedding layer that converts the words into lower dimensional internal representation vectors, the LSTM layer, and an output layer consisting of a single neuron with a sigmoid activation that outputs a value indicating if the review is positive ($y \geq 0.5$) or negative ($y < 0.5$).

The IMDb dataset was obtained using the \textsc{Keras} \cite{keras} \texttt{datasets} application programming interface (API), with the preprocessing done with custom code \footnote{\url{https://github.com/NicholasCJL/imdb-LSTM/blob/master/data_processing.py}}. The networks were trained using \textsc{Keras} with the \textsc{TensorFlow 2.6.0} \cite{tensorflow} backend on a GeForce GTX 1080 GPU, with preprocessing steps performed on a Ryzen 9 3900X CPU. 



The hyperparameters used for the LSTM networks are shown in \autoref{tab:hyperparam}. These hyperparameters were selected with the \textsc{KerasTuner} \cite{kerastuner} library using the \texttt{Hyperband} \cite{hyperband} search algorithm, selected over 10 hyperband iterations. Overall, we follow the best practices of the state-of-the-art for LSTM models in this work.

\subsection{\label{sec:expt-measurement}Measuring $1/f$ noise}

In order to measure the exponent of the power spectrum in the LSTM cell, temporal sequences of the specific activations have to be obtained. To obtain the internal activations of the \textsc{Keras} LSTM cells, the cell was recreated in vanilla Python with \textsc{NumPy} \cite{numpy}. The code for this is available at \footnote{\url{https://github.com/NicholasCJL/imdb-LSTM/blob/master/get_LSTM_internal_vectorized.py}}. Detrended Fluctuation Analysis (DFA) \cite{DFA:original} is then used for analysis of the time series to obtain the scaling exponent of the series. In this work, we use a modified version of the algorithm, the unbiased DFA \cite{DFA:unbiased} in order to account for the relatively short time series due to the length 500 reviews.

The steps to obtain the exponent of any specific activation ($\boldsymbol{f}$ in this example) is then as follows:
        
\begin{enumerate}
    \item Propagate the review through the LSTM layer, recording the vector $\boldsymbol{f}_t$ corresponding to the forget gate of each LSTM cell at each timestep $t$, forming 8 time series of activations corresponding to the 8 LSTM cells.
    \item Sum the activations across the cells in the layer to get the total activation $S(t)$ of the LSTM layer.
    \item Calculate the fluctuation $F(t)$ of the activation with DFA.
    \item Calculate the slope of the log-log plot of $F(t)$ vs $t$ to obtain the DFA scaling exponent $\alpha$.
    \item Calculate the power spectrum exponent $\beta = 2\alpha - 1$.
\end{enumerate}

\section{\label{sec:results}Results and Discussion}

When picking the optimal epoch of the networks, the epoch with the lowest network loss was selected. The accuracies achieved across the networks were high \footnote{For reference, the best-performing LSTM based classifier has a 94.1\% accuracy, and the best-performing model is a transformer with 96.1\% accuracy. Taken from \url{https://paperswithcode.com/sota/sentiment-analysis-on-imdb}}, ranging from 87.61\% to 89.63\% prediction accuracy on the test data. 



\begin{figure}[!h]
    \subfloat[]{
        \includegraphics[width=\linewidth]{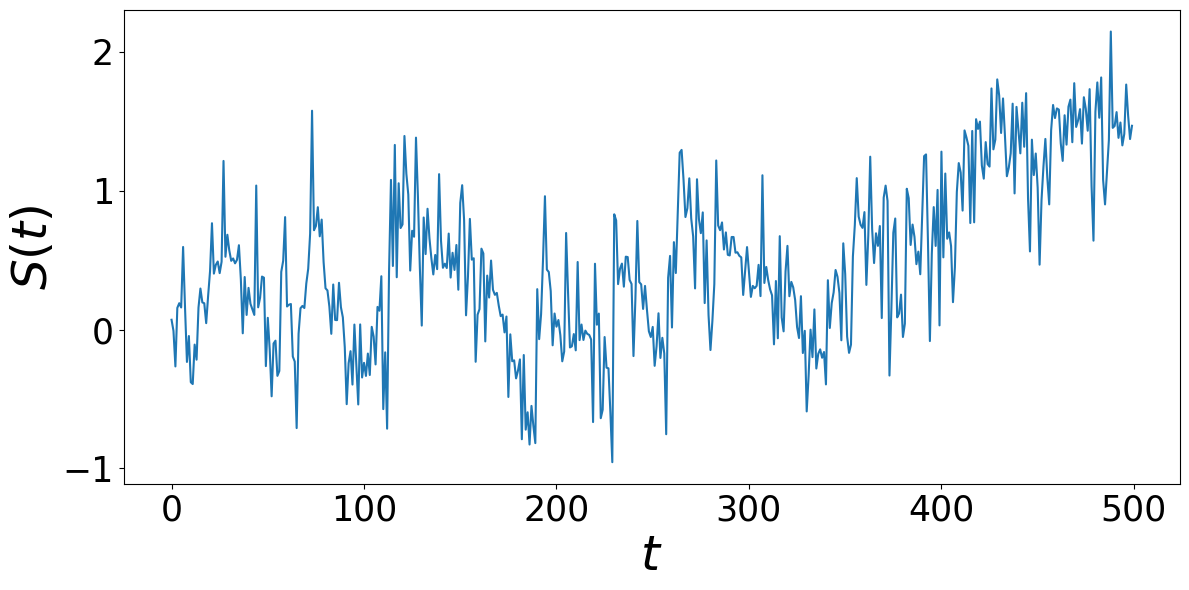}
    }
    \hspace{1em}
    \subfloat[]{
        \includegraphics[width=\linewidth]{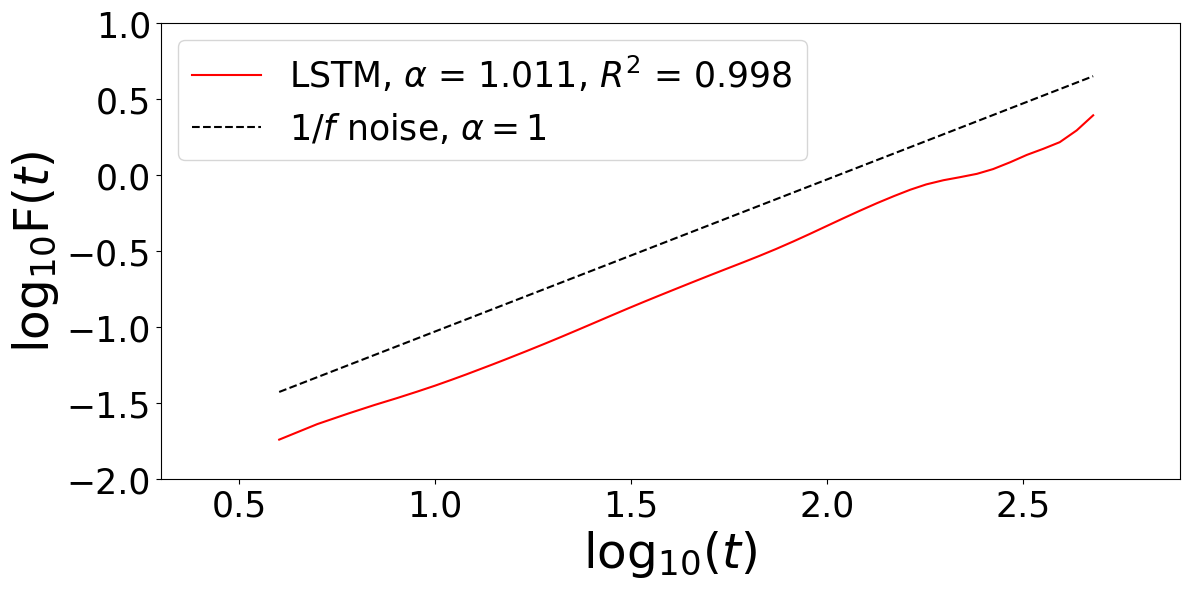}
    }
    \hspace{1em}
    \subfloat[]{
        \includegraphics[width=\linewidth]{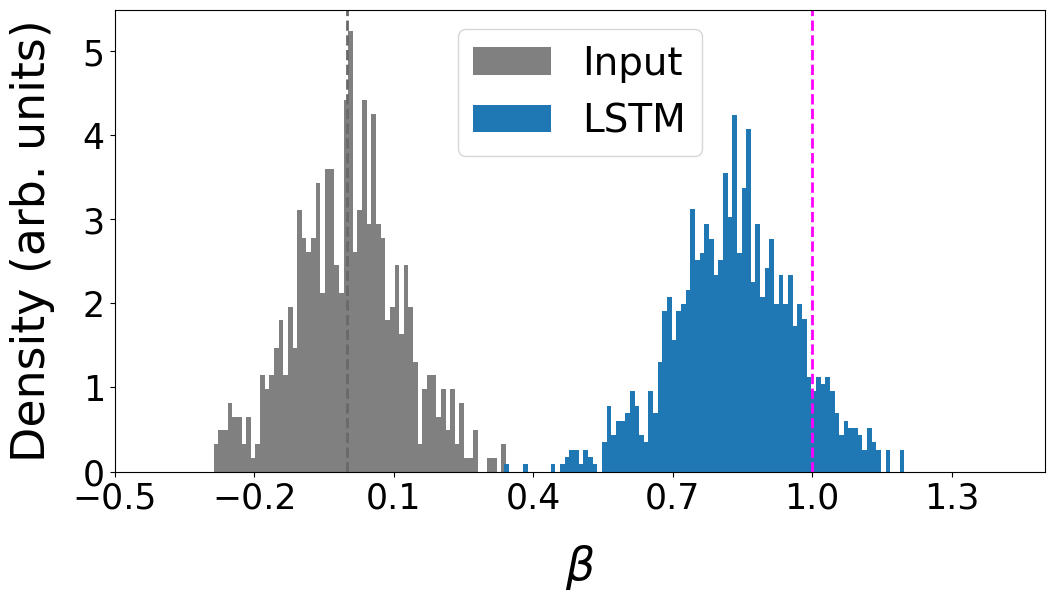}
    }
    \caption{\label{fig:resultsbig-a}(a) Value of the activation $\boldsymbol{h}_t$ for the entire LSTM layer for a single 901 word review (truncated to 500 words). \label{fig:resultsbig-b}(b) Log-log plot of the fluctuation $F(t)$ (solid, red) of the activation shown in (a) obtained with DFA with a reference line for $1/f$ noise (dashed, black) plotted. \label{fig:resultsbig-c}(c) Histogram of the exponents $\beta$ obtained from the DFA exponents $\alpha$ across all the test reviews with length $\ge$ 500. Exponents for the input embedding activations (grey) and $\boldsymbol{h}_t$ (blue) are both plotted. Dotted lines corresponding to $\beta=0$ and $\beta=1$ plotted for reference.}
\end{figure}

\subsection{\label{sec:results-exponent}Exponent $\beta$ for the test set}

The steps described in \autoref{sec:expt-measurement} were performed for the reviews in the test dataset of length $\ge 500$ to remove the impact of the padding as the repeated identical padding tokens has the effect of lowering the exponent. Note that training of the LSTM is carried out on reviews regardless of their word lengths, to keep in line with the accepted practices in AI. \autoref{fig:resultsbig-a}(a) shows the activations of $\boldsymbol{h}$ for one review for one of the networks. \autoref{fig:resultsbig-b}(b) is a log-log plot of the fluctuation $F(t)$ against time, with a reference line with gradient 1 representing $\beta=2\alpha-1=1$ shown. \autoref{fig:resultsbig-c}(c) displays a histogram of the exponents of $\boldsymbol{h}$ obtained for all the test reviews with length $\ge 500$ for the same network with a mean $\mu_\beta = 0.833 \pm 0.132$.

\begin{figure}[h]   
    \includegraphics[width=\linewidth]{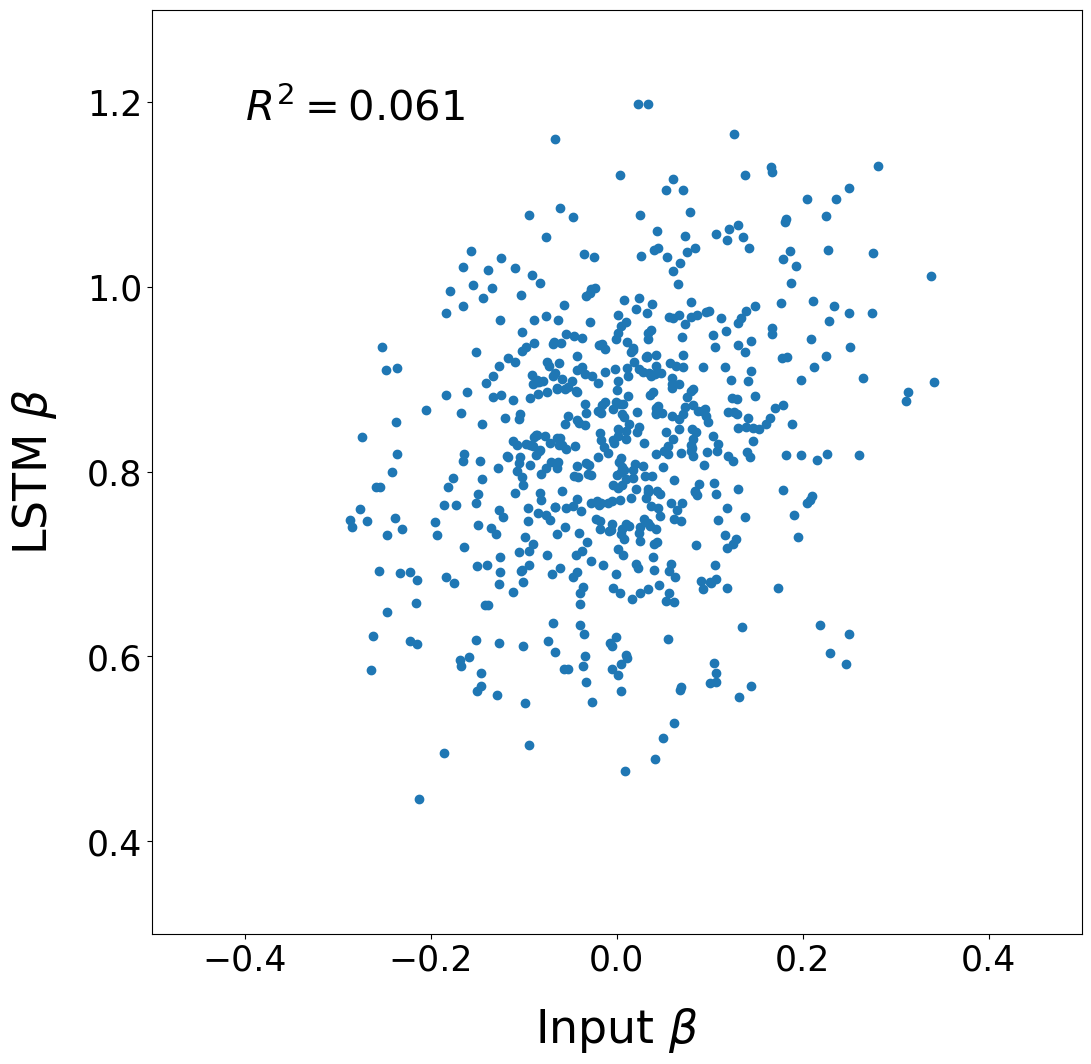}
    \caption{\label{fig:inputvactivations}Scatter plot of the exponents of $\boldsymbol{h}_t$ vs the exponents of the input $\boldsymbol{x}_t$ for test reviews with length $\ge 500$. Model used is the same as \autoref{fig:resultsbig-a}. The input exponents here are not correlated to the activation exponents, showing that the $1/f$ phenomenon in the activation values is not from a similar pattern in the inputs.}
\end{figure}

\subsection{\label{sec:results-input}Ruling out $1/f$ noise in the input}

One possibility for the presence of $1/f$ noise is that the input $\boldsymbol{x}$ has a $1/f$ signal. As such, it is important to rule out this effect if we were to demonstrate the emergence of $1/f$ noise from the LSTM. To determine the exponent of the input data, the same process from \ref{sec:expt-measurement} was performed, using the embedding vector instead of the activation vector.

 The distribution of exponents of the inputs to the LSTM (after embedding) is also plotted in \autoref{fig:resultsbig-c}, showing effectively uncorrelated white noise with a mean $\mu_\beta = 0.002 \pm 0.116$. 

\autoref{fig:inputvactivations} is a scatter plot relating the histograms shown in \autoref{fig:resultsbig-c}, demonstrating the lack of correlation between the activation exponent and the input exponent with an $R^2$ value of 0.061. This further supports our hypothesis that the $1/f$ noise observed in the LSTM networks are inherent to the networks, rather than a consequence of $1/f$ noise in the inputs to the networks.

\subsection{\label{sec:results-conditions}Conditions for $1/f$ noise}

\begin{table*}[ht]
    \centering
    \caption{\label{tab:acc}Average accuracies for the LSTM networks of different sizes}
    \begin{ruledtabular}
    \begin{tabular}{cccccccccc}
        $n$           & 8     & 16    & 32    & 48    & 64    & 80    & 96    & 112   & 128   \\
        \colrule
        Mean Accuracy (\%) & 88.95 & 88.97 & 88.78 & 88.79 & 88.96 & 88.56 & 88.32 & 88.78 & 88.24 \\
    \end{tabular}
    \end{ruledtabular}
\end{table*}

\begin{figure}  
    \includegraphics[width=\linewidth]{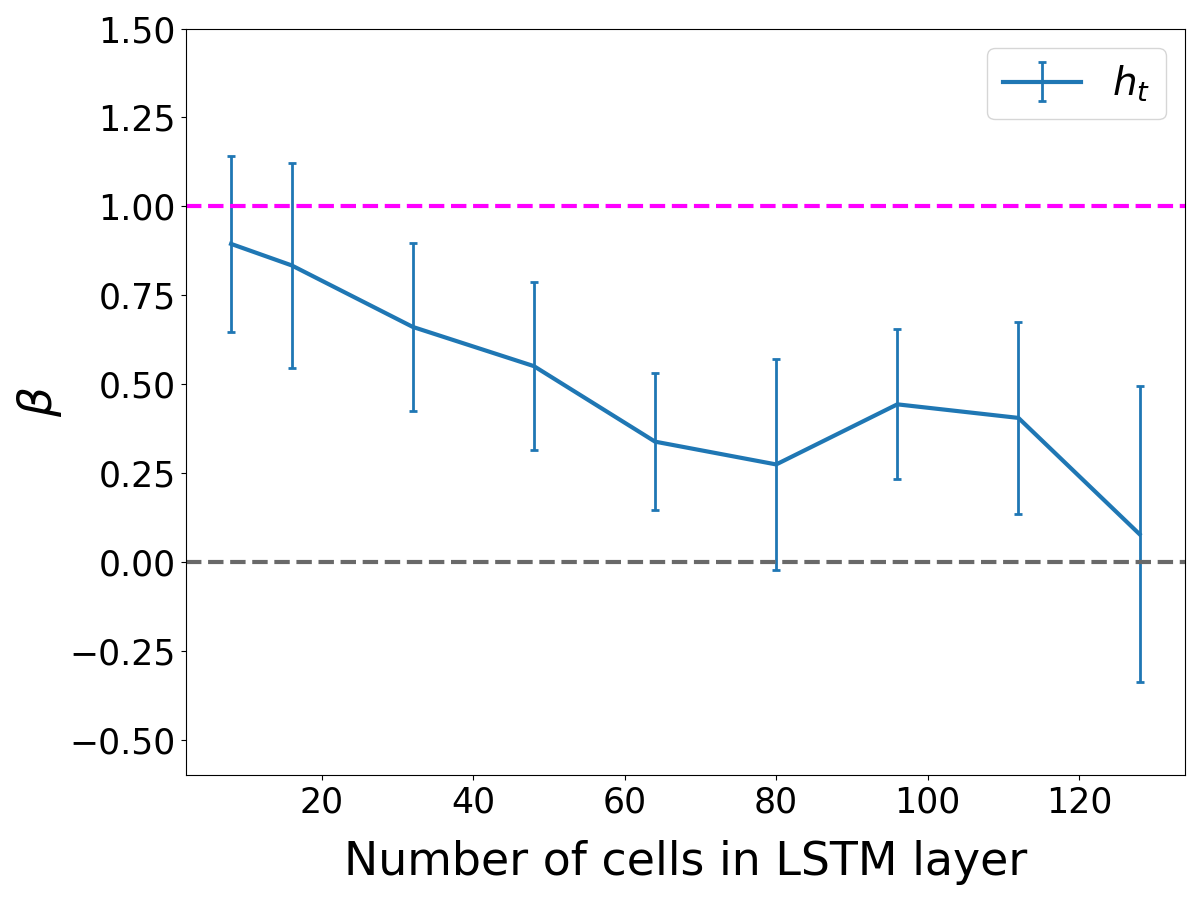}
    \caption{\label{fig:ht_trend}Aggregate values of $\beta$ for the activation $\boldsymbol{h}_t$ vs the number of cells in the LSTM layer. $1/f$ noise (dashed, pink) and white noise (dashed, black) are shown for reference.}
\end{figure}

Drawing inspiration from the analogy to the brain and the presence of $1/f$ noise in a well-functioning brain, we investigated the exponents of the LSTM layer while varying the capacity of the layer. The greater the number of cells in the LSTM layer, the greater its capacity for storing useful information. \autoref{fig:ht_trend} shows the aggregate values of $\beta$ across LSTM networks of different cell sizes $n$, aggregated over multiple LSTM networks of the same cell size trained on the same dataset with different random initialisations. We observe a decreasing trend for the exponent $\beta$ as the network capacity increases. At a size of 128 cells, the aggregate activation of the LSTM layer effectively approaches white noise. Despite the decreasing trend in $\beta$, the average accuracies of the different networks are approximately constant, demonstrated in \autoref{tab:acc}. Appendix \ref{app:other_trends} contains data for the other activations within the LSTM layer, with the other activations demonstrating a similar trend.

\begin{figure}[!h]
    \subfloat[]{
        \includegraphics[width=\linewidth]{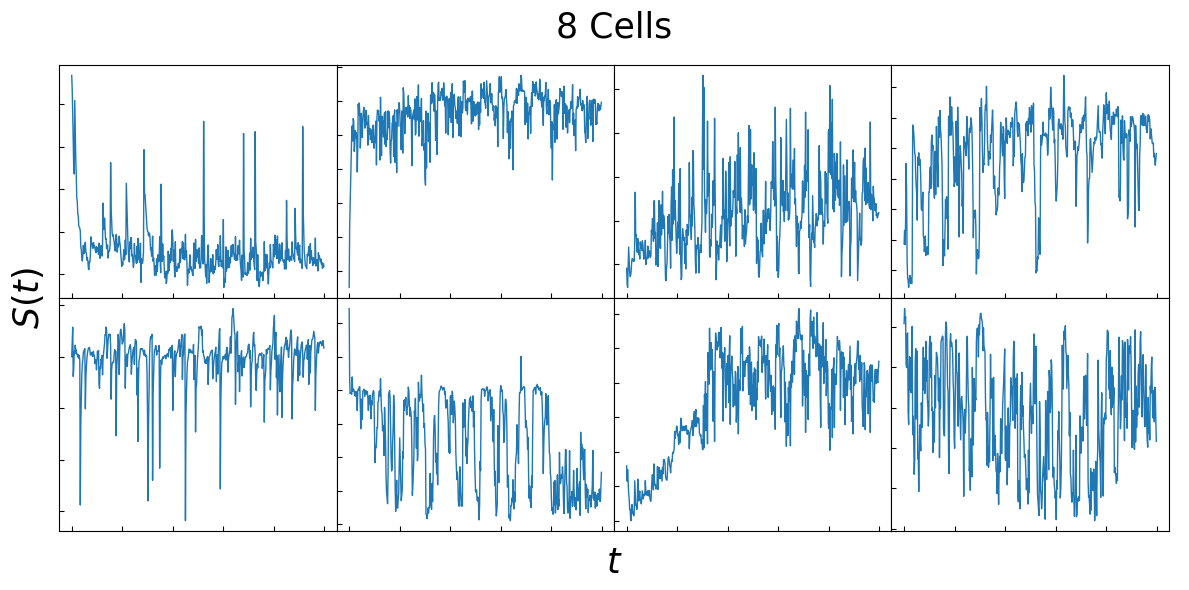}
    }
    \hspace{1em}
    \subfloat[]{
        \includegraphics[width=\linewidth]{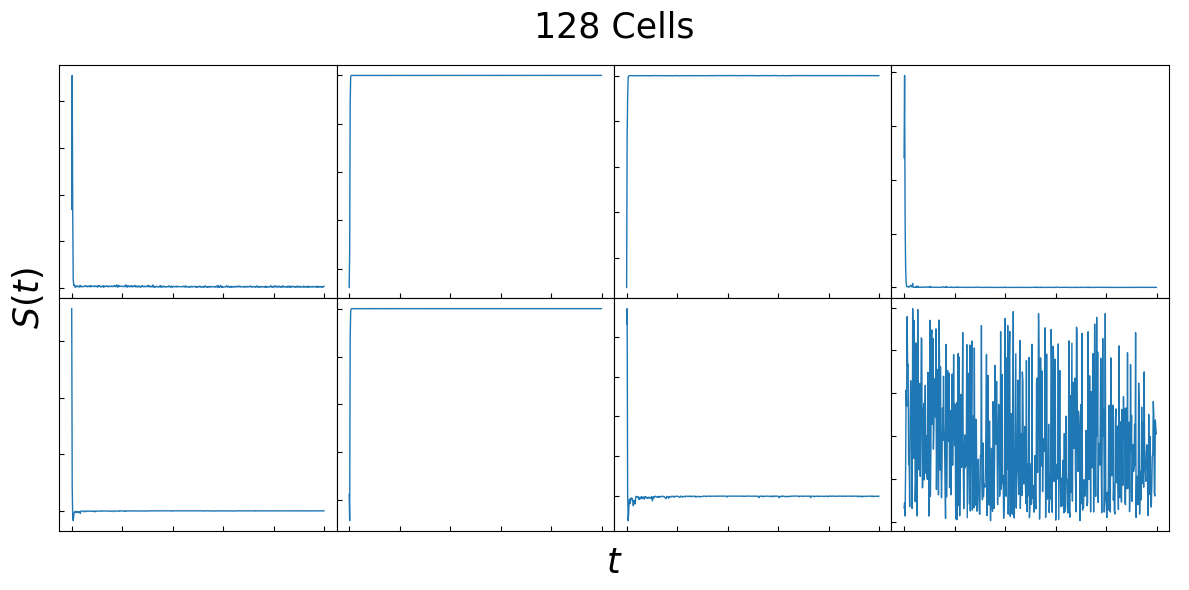}
    }
    \caption{\label{fig:cell-level-comparison}Cell-level activations of $\boldsymbol{h}_t$ for the same review in an LSTM layer of 8 cells (a) and 128 cells (b). The first 8 cells are shown here for the 128 cell layer. We observe here the lack of activity in multiple cells for the large LSTM layer.}
\end{figure}

A possible reason for this can be seen when analysing the individual cells of the LSTM layers. \autoref{fig:cell-level-comparison} shows the activation $\boldsymbol{h}_t$ for the first 8 cells of the smallest LSTM layer (8 cells) and the largest LSTM layer (128 cells) processing the same review. We see a clear indication of multiple cells in the large layer being ``dead'', with little to no activity in the cells. Appendix \ref{app:cell-level} shows the same plots for the LSTM layers of other sizes with the same review, demonstrating the decreasing activity of the layers as their capacity increases.

\begin{figure}  
    \includegraphics[width=\linewidth]{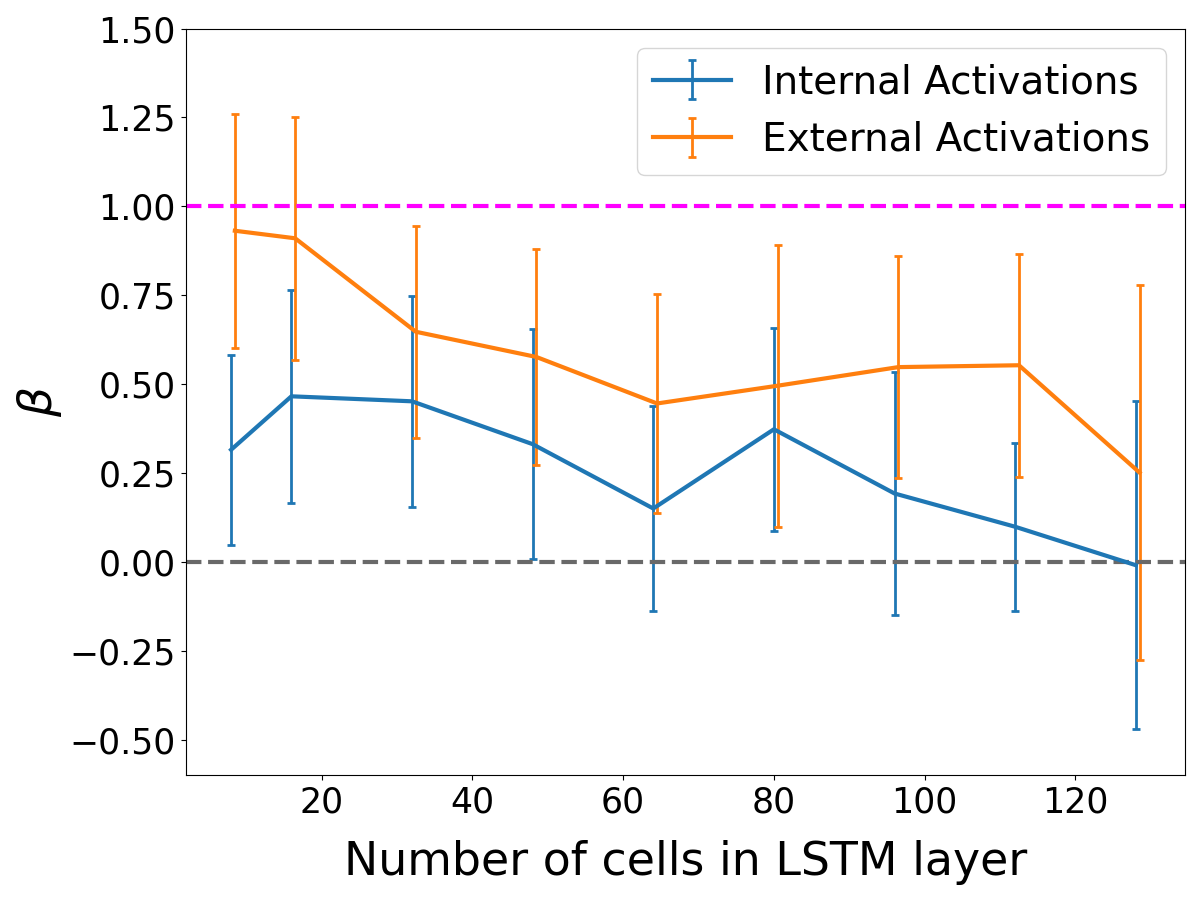}
    \caption{\label{fig:internal_v_external}Aggregate values of $\beta$ for external activations ($\boldsymbol{h}_t$, $\boldsymbol{c}_\mathrm{out}$) and internal activations ($\boldsymbol{i}_t$, $\boldsymbol{f}_t$, $\boldsymbol{\tilde{c}}_t$, $\boldsymbol{o}_t$) vs number of cells in the LSTM layer. Line for external activations displaced horizontally slightly to increase visibility of error bars. $1/f$ noise (dashed, pink) and white noise (dashed, black) are shown for reference.}
\end{figure}

\subsection{\label{sec:internal-v-external}``Internal'' and ``External'' activations}

In the measurements of $1/f$ noise in the human brain, the EEG measurements measure the neurons on the surface of the brain, while the fMRI measurements measure the neurons in the volume of the brain. These two measurements show markedly different exponents in their activations with the bulk exponents being lower than the surface exponents, though the values of $\beta$ in both types of measurements are close to 1. In our experiments, we observed a similar split in the values of the exponents in what we term ``internal'' activations -- which serve to regulate the layer output -- and ``external'' activations -- which contain the values of the layer output.

\autoref{fig:internal_v_external} contains a plot of $\beta$ across the same LSTM networks of different sizes, with the aggregate values of the internal and external activations shown. We observe in both cases the same trend seen previously, with the value of $\beta$ decreasing as network capacity increases. However, we also see a clear split in $\beta$ between both types of activations, with the internal activations consistently having a smaller $\beta$ than the external activations. Such similar patterns to the brain signals is probably more than just coincidence, and may need further investigation.



\section{\label{sec:conclusion}Conclusion}

In this study, we have found convincing evidence of $1/f$ noise in deep neural networks like the Long-Short-Term Memory (LSTM) network. This pattern has been demonstrated to be independent of the input data which exhibits white noise. Since the input data consists of real world natural language sentences that our brain processes, our results demonstrate that artificial neural networks with close to human-level performance exhibit very similar $1/f$ patterns as their biological counterparts\cite{noise:meg, brain:overview, fMRI:noise}.
Interestingly, the analogy also applies to the trends in the noise exponents for ``internal" and ``external" neurons within the LSTM compared to fMRI and EEG exponents respectively \cite{EEGMEG:noise, fMRI:noise}. Internal activations generally have a smaller exponent $\beta$ in the power-law spectrum $f^{-\beta}$. 

However, it is worth noting that the $1/f$ noise pattern fades away when the capacity of the neural network increases to overcapacity for the specific learning task. Specifically, when we increase the number of neurons without observing a clear improvement in model accuracy, we find that many neurons do not respond to input fluctuations, effectively not participating in ``thinking'' for the task, instead demonstrating low to no activity. The abundance of these unactivated neurons effectively reduces the exponent $\beta$ of the aggregate neuron fluctuations.

It is intriguing that despite the vast differences in the microscopic details between biological neural networks and artificial neural networks, such macroscopic pattern of the $1/f$ noise are strikingly similar. This similarity points at some deeper principles that govern their healthy functioning, something that is seemingly independent of the detailed neural interactions since the brain neural network is drastically different from LSTM. With artificial neural networks being more ``transparent'' to our experimental manipulation and examination unlike their biological counterparts, they are ideal proxies to understand the origin of $1/f$ noise going forward, as well as possible tools to understand more about the healthy functioning of the brain through the $1/f$ noise perspective.

Overall, our findings open up new interdisciplinary avenues for exploring the operational regimes and computational principles governing both biological and artificial neural networks grounded upon 1/f noise.


\section{Appendixes}
\appendix
\section{Exponent vs Number of Cells (Other Activations)\label{app:other_trends}}

The following figures display the same trend seen in Figure \ref{fig:ht_trend} for the other activations in the LSTM layer. Error bars shown represent 1 standard deviation, $1/f$ noise (dashed, pink) and white noise (dashed, black) are shown for reference.

\begin{figure}[H] 
    \includegraphics[width=\linewidth]{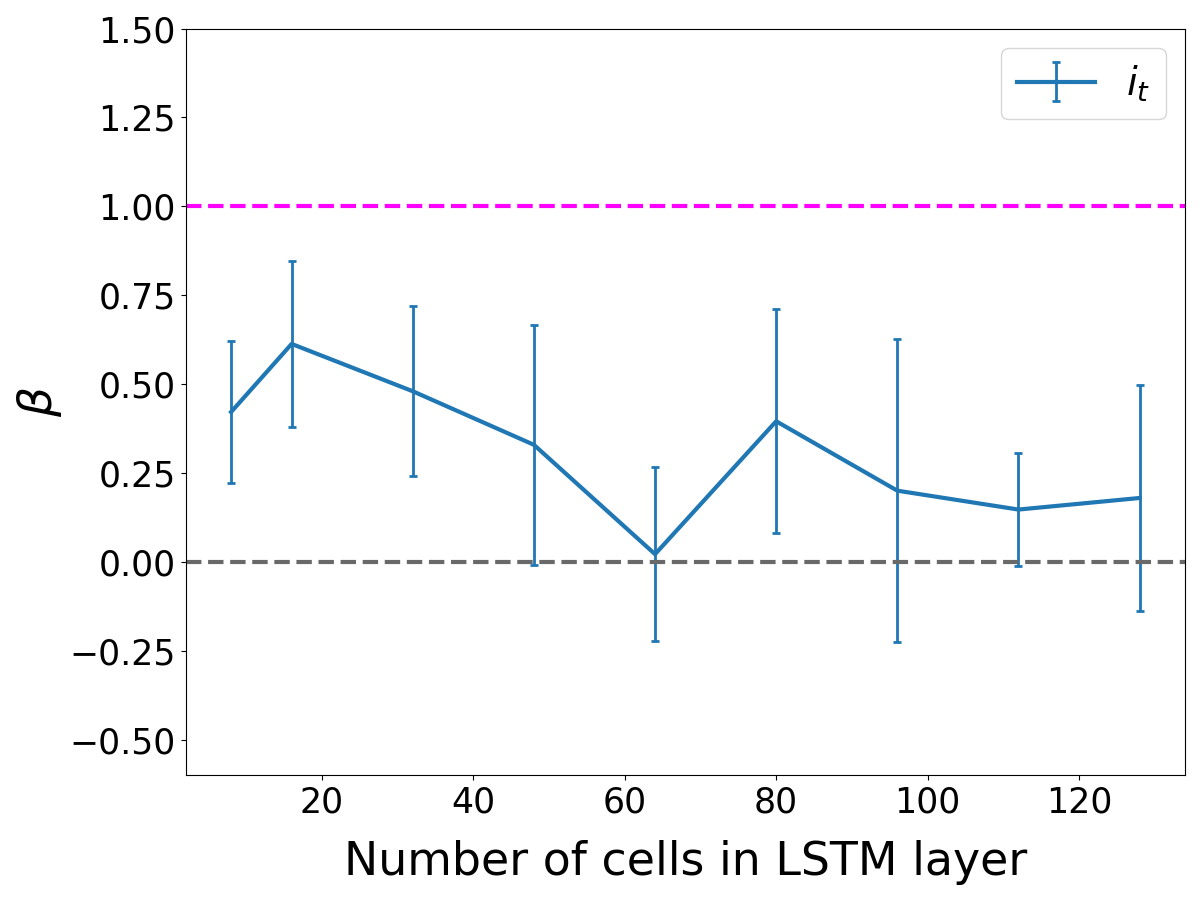}
\end{figure}

\begin{figure}[H]
    \includegraphics[width=\linewidth]{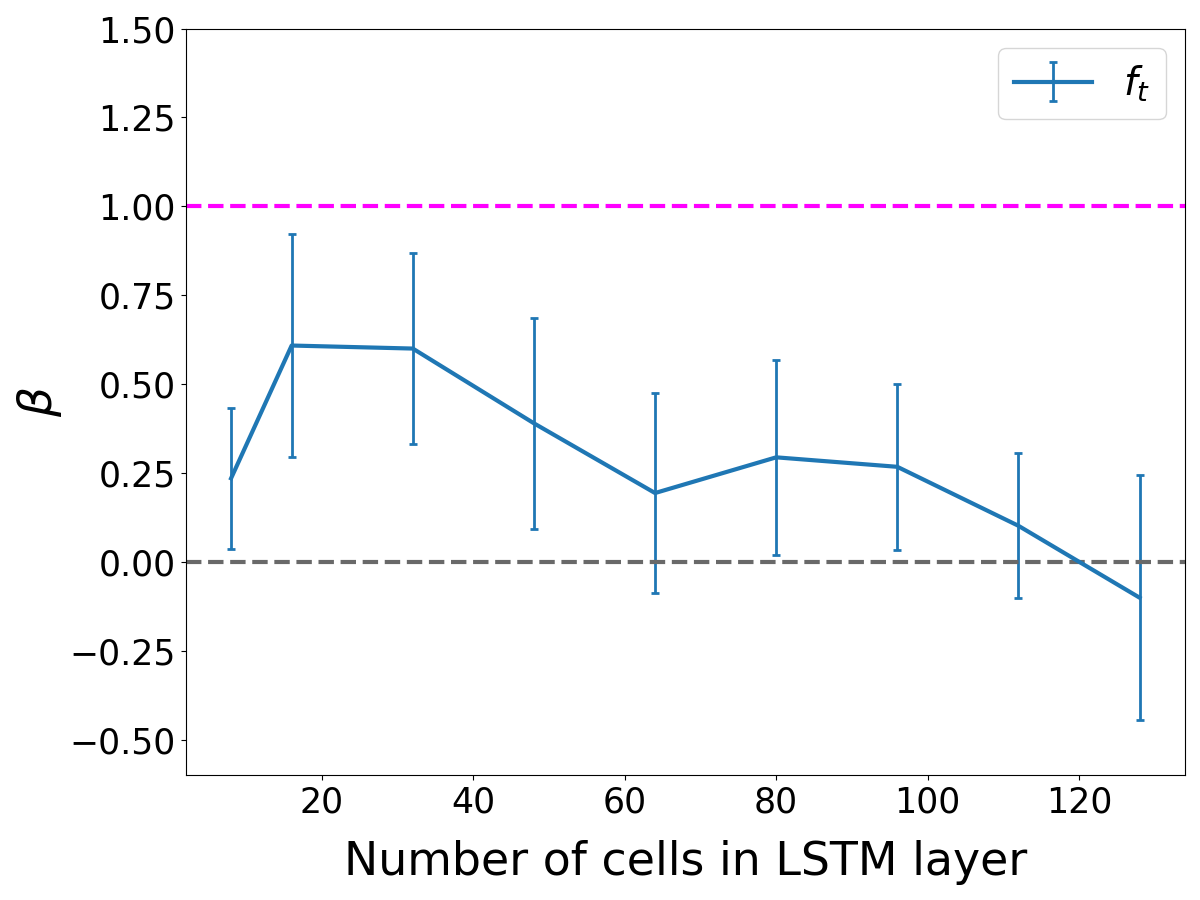}
\end{figure}

\begin{figure}[H]
    \includegraphics[width=\linewidth]{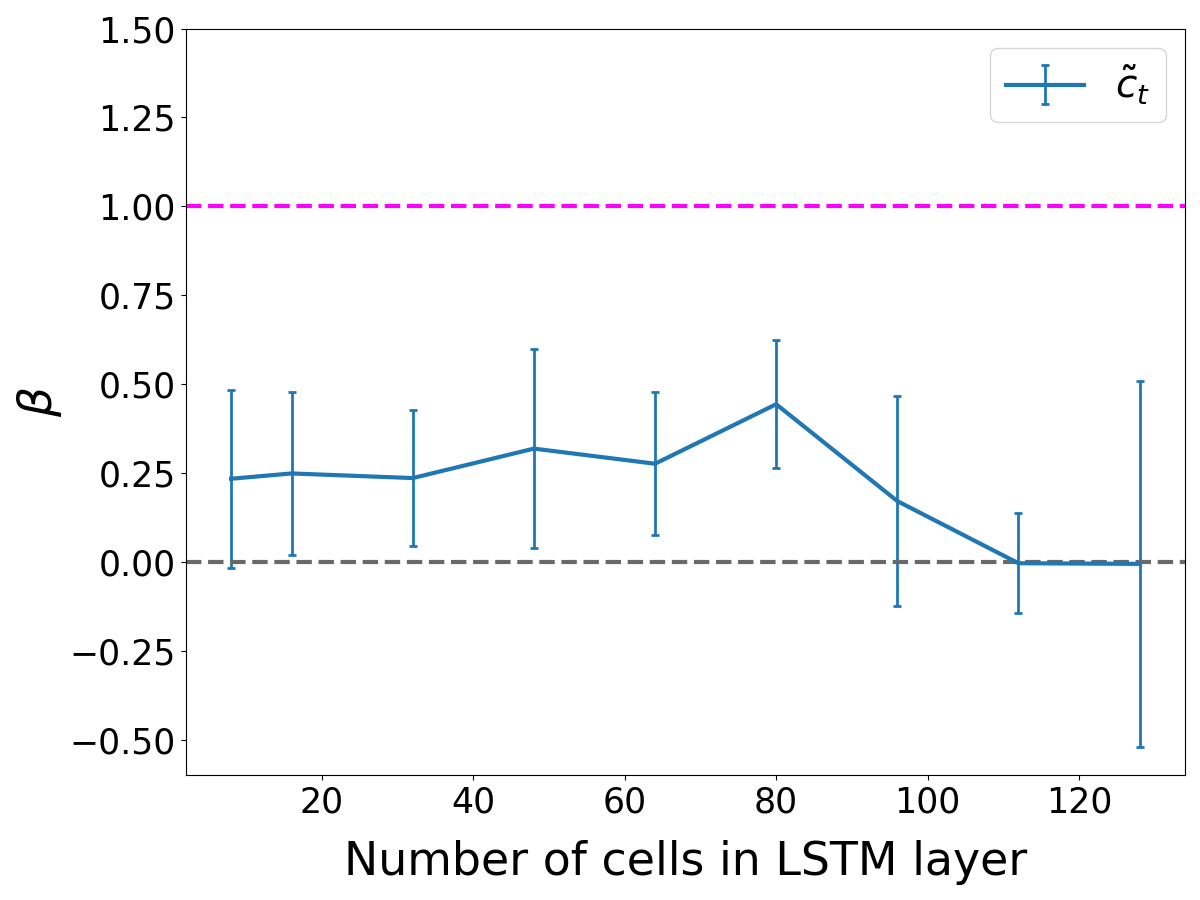}
\end{figure}

\begin{figure}[H]
    \includegraphics[width=\linewidth]{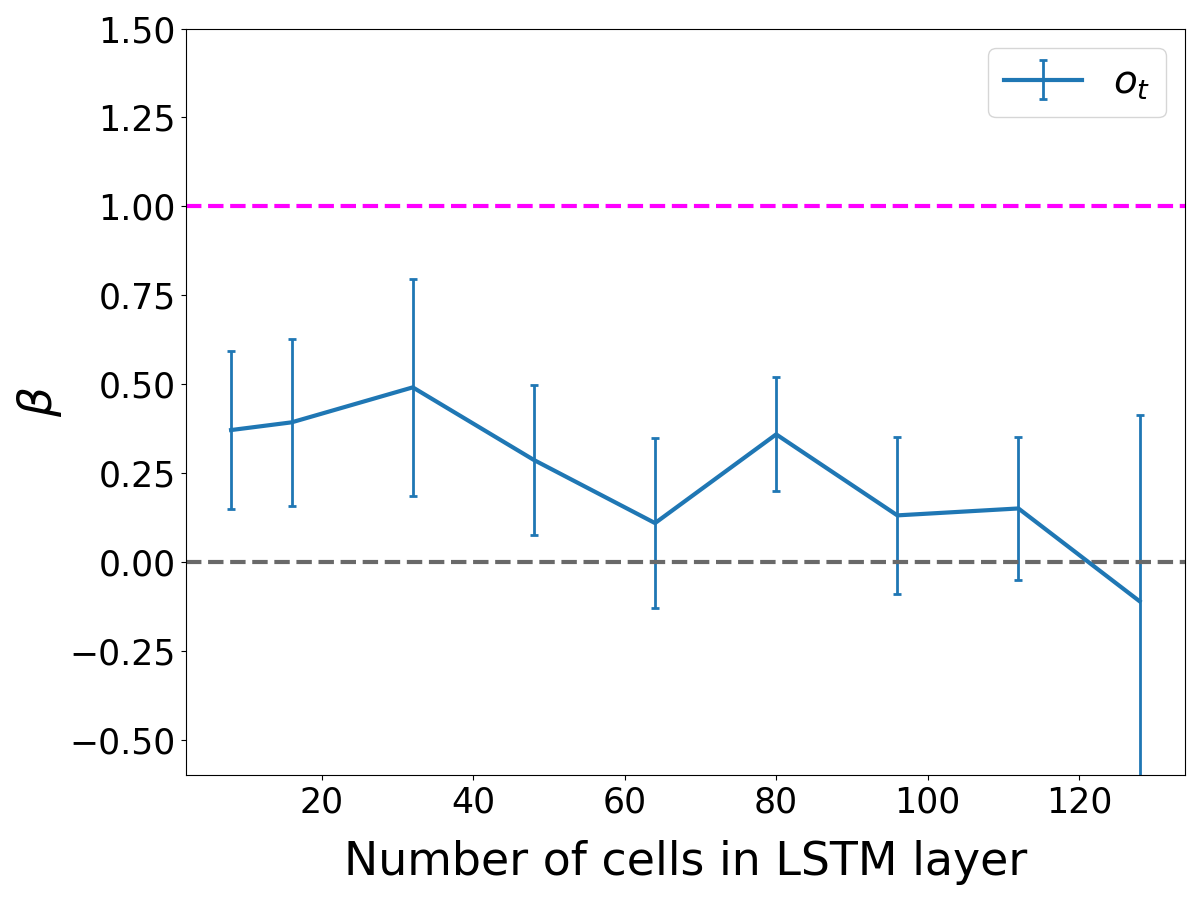}
\end{figure}

\begin{figure}[H]
    \includegraphics[width=\linewidth]{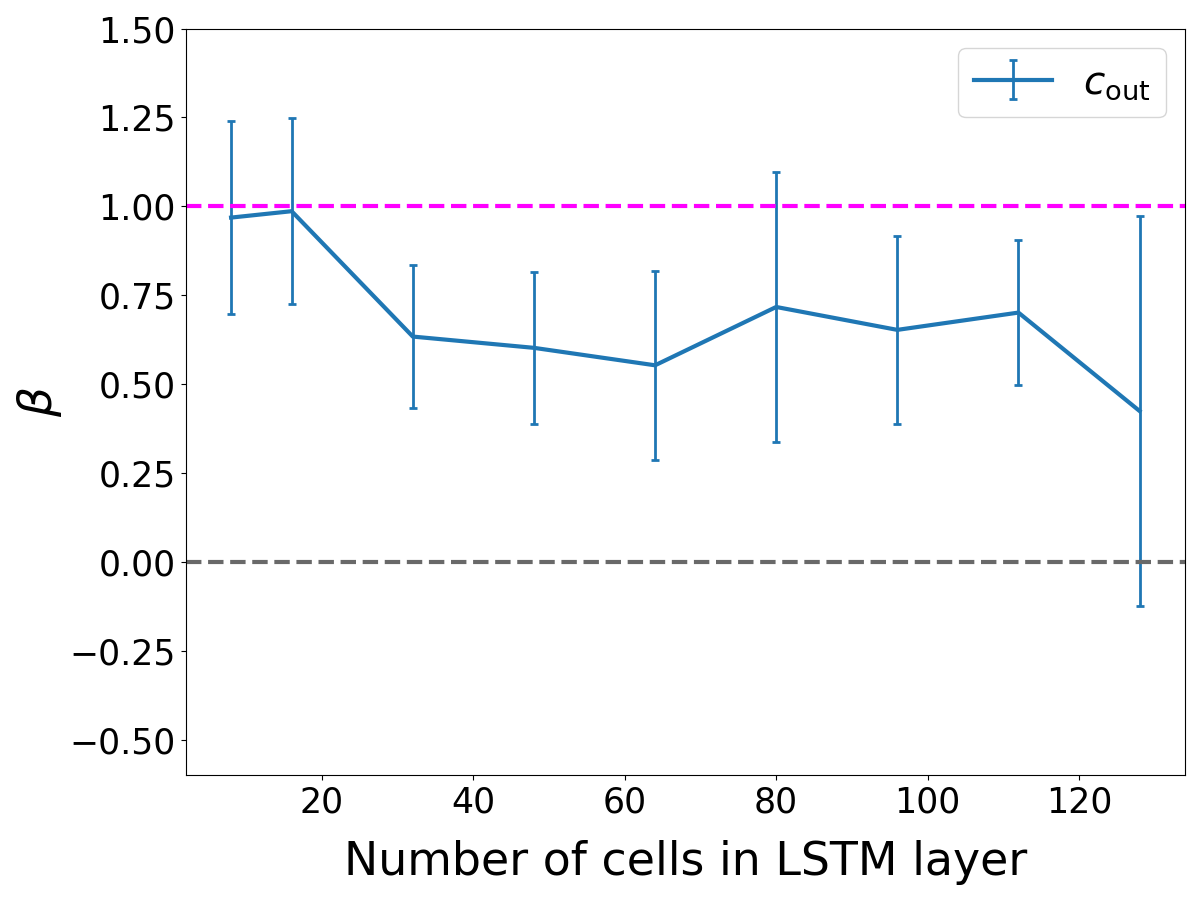}
\end{figure}

\section{Cell-Level Activations\label{app:cell-level}}

The following figures display the cell-level activations for the LSTM layers of different sizes, similar to in Figure \ref{fig:cell-level-comparison}. Each figure shows the activation $\boldsymbol{h}_t$ for the 8 of the cells within the LSTM layer. The same review is supplied to the LSTM layers across all the figures. We observe the previously described effect of loss of activity in some cells as the size of the LSTM layer increases.

\begin{figure}[H]
    \centering
    \includegraphics[width=\linewidth]{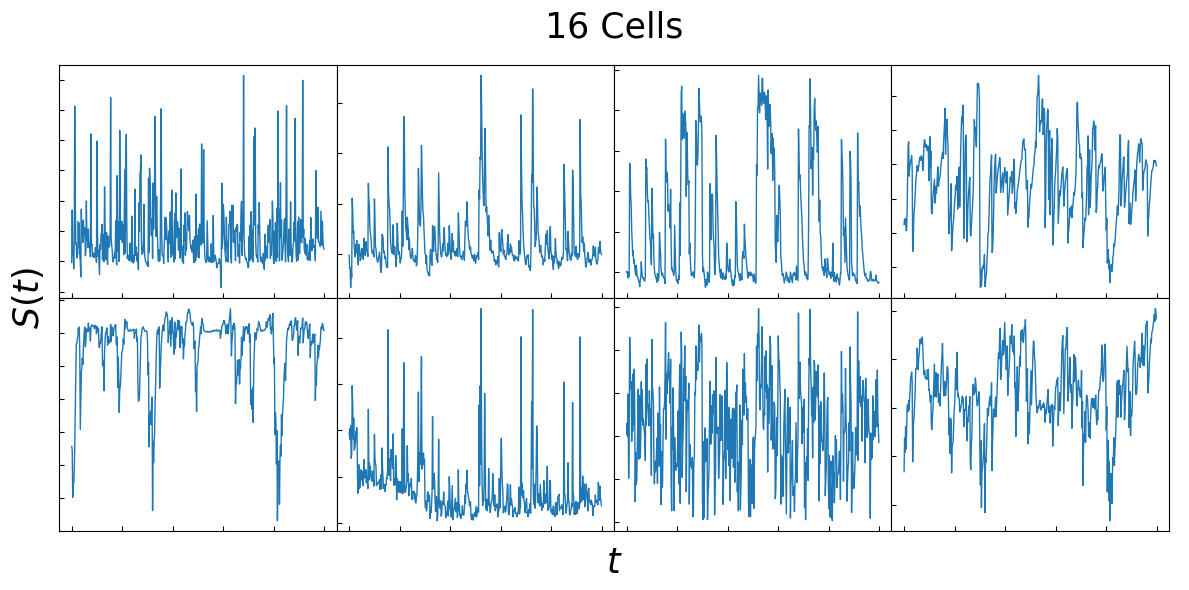}
\end{figure}

\begin{figure}[H]
    \centering
    \includegraphics[width=\linewidth]{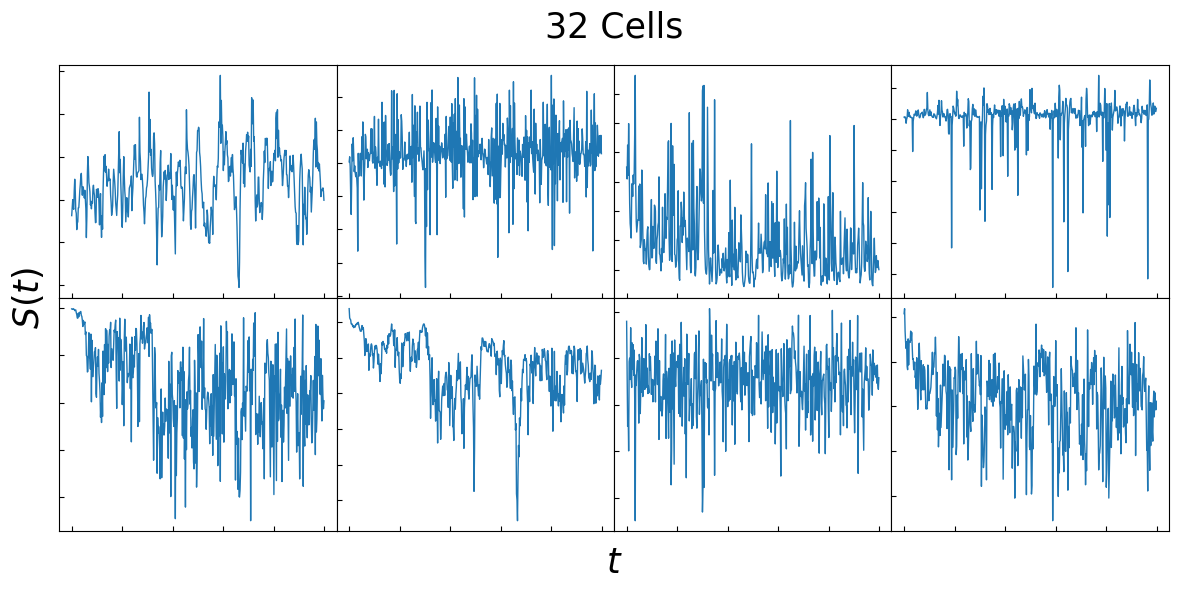}
\end{figure}

\begin{figure}[H]
    \centering
    \includegraphics[width=\linewidth]{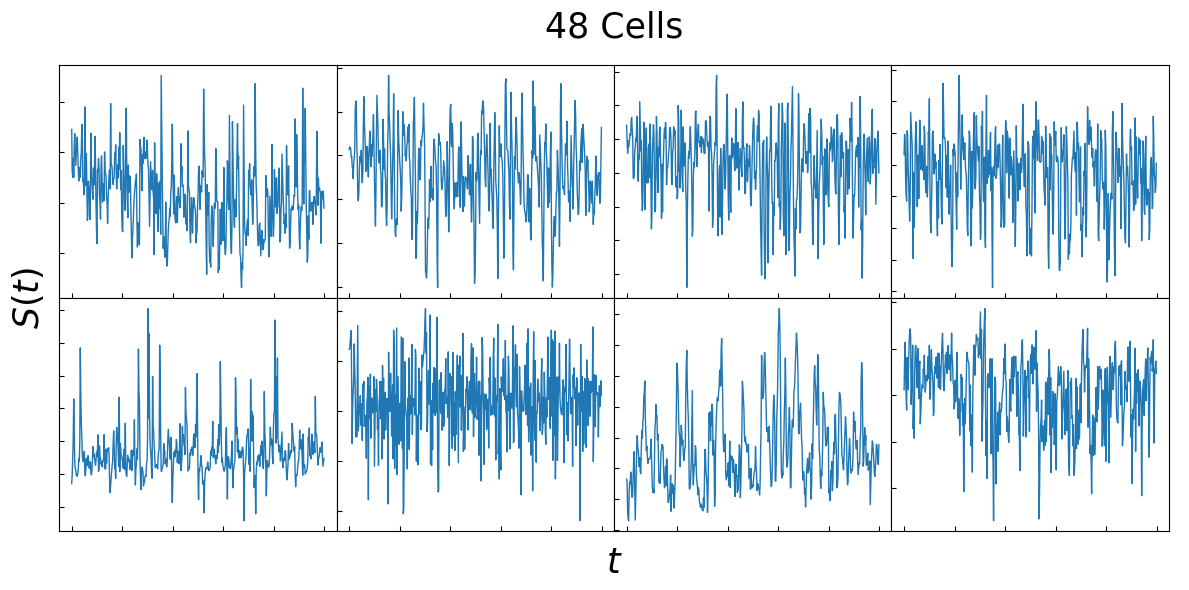}
\end{figure}

\begin{figure}[H]
    \centering
    \includegraphics[width=\linewidth]{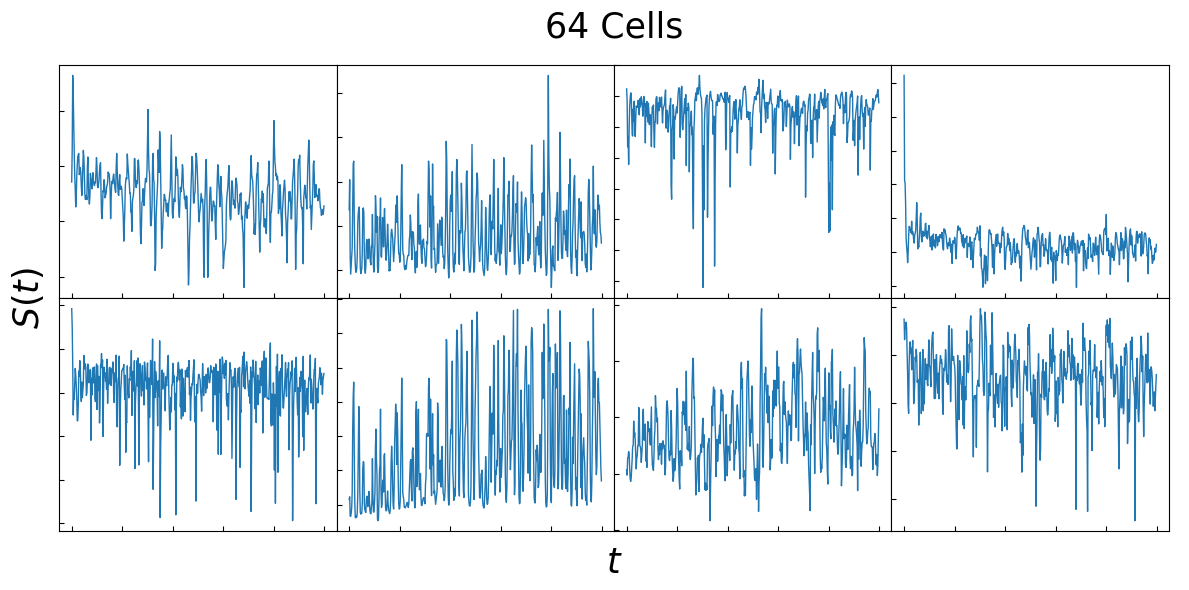}
\end{figure}

\begin{figure}[H]
    \centering
    \includegraphics[width=\linewidth]{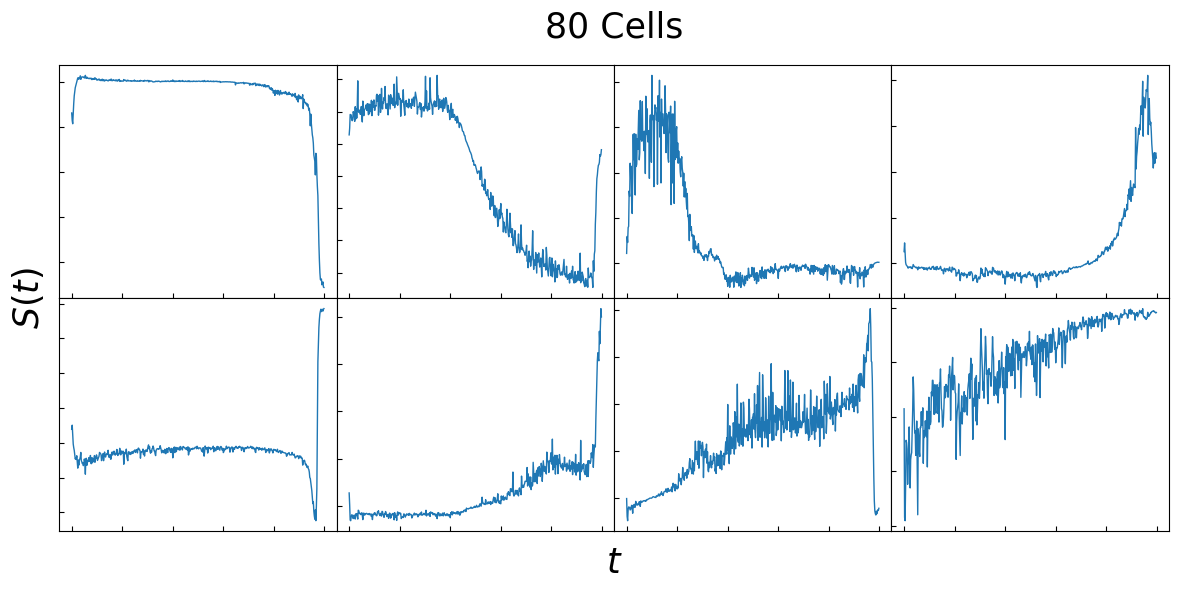}
\end{figure}

\begin{figure}[H]
    \centering
    \includegraphics[width=\linewidth]{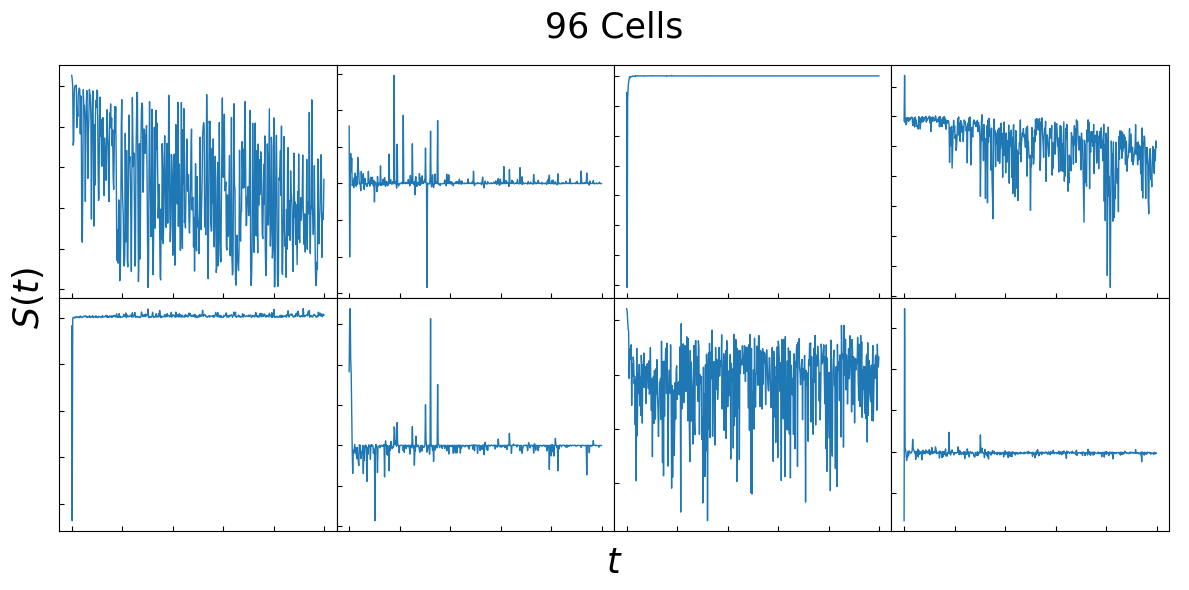}
\end{figure}

\begin{figure}[H]
    \centering
    \includegraphics[width=\linewidth]{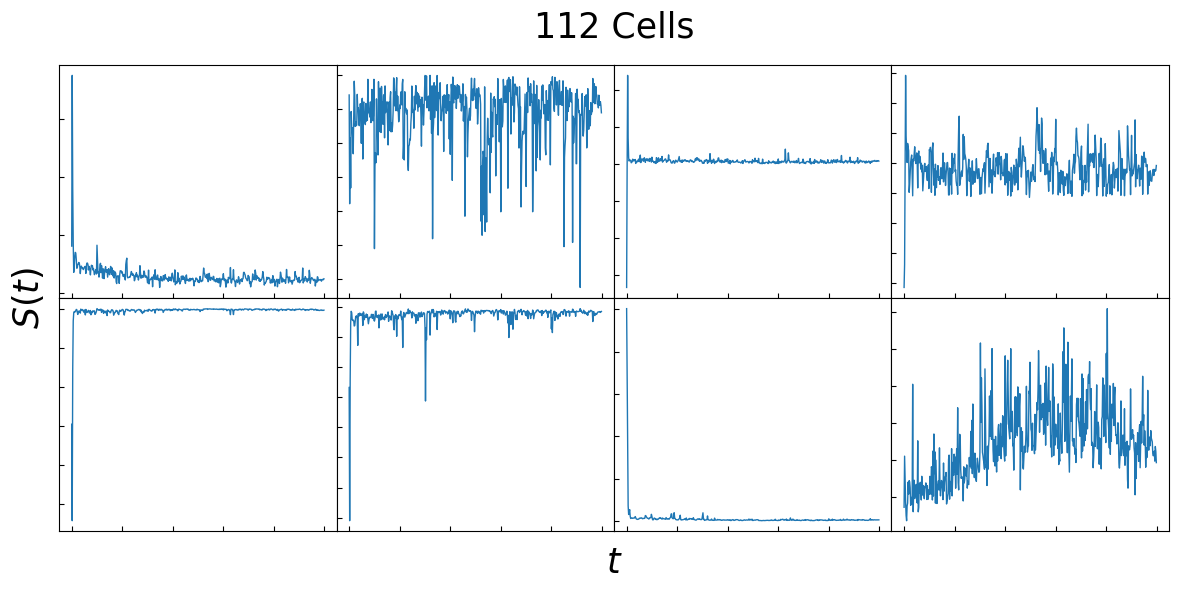}
\end{figure}


\bibliography{apssamp}

\end{document}